# The secret chambers in the Chephren Pyramid

Bartosz Gutowski[1], Witold Jóźwiak[1], Markus Joos[2], Janusz Kempa[3], Kamila Komorowska[1], Kamil Krakowski[1], Ewa Pijus[1], Kamil Szymczak[1], Małgorzata Trojanowska[1]

1. L.O. im. Marsz. Małachowskiego, Płock
2. CERN
3. The Institute of Civil Engineering, Faculty of Civil Engineering, Mechanics and Petrochemistry, Warsaw University of Technology

# Contents

# Introduction: the mystery of the pyramids

Pyramids are the greatest architectural achievement of ancient civilization, so people all over the world are curious as to the purpose of such huge constructions. No other structure has been studied as thoroughly, nor have so many books and articles been written about it. However, the pyramids are still the most mysterious pieces of architecture on earth and it seems that they still hide numerous mysteries. One of the biggest and the most intriguing to us is the internal structure of the pyramid of Chephren (Khafre). The Chephren Pyramid does not seem to hide any more than a modest chamber (known as the "Belzoni Chamber"), unlike his father's, Cheops', Great Pyramid, which has many sophisticated internal structures: the King's Chamber, the Queen's Chamber and the Great Gallery. The fact that the Egyptians set up 2.5 million stone blocks without any purpose seems to be unimaginable. Therefore, we attempt to examine the internal structure of the pyramid using muon tomography [1].

In the 1930s, the attenuation of muons under the ground was used to measure the depth of the London subway [2]. In recent years, Japanese teams have used muon tomography to estimate the level of lava in volcanoes [3] and to check the damages of the nuclear reactor in Fukushima after the tsunami of 2011 [4].

# A mathematical model and calculations

We created a computer model of the pyramid. To validate the model, we compare our calculations with the experimental data of Luis W. Alvarez [5]. L. Alvarez installed his detectors (spark chambers) inside the Belzoni Chamber. This chamber is displaced from the centre of the base by $y_0$ =13.5 m to the East and $x_0$ = 4 m to the North (Figure 2).

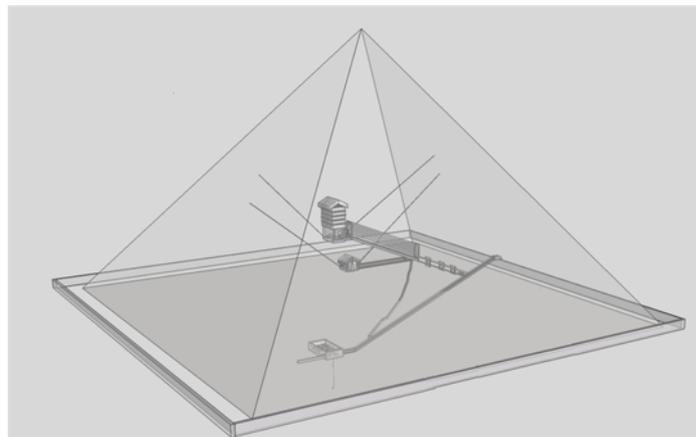

Figure 1: 3D representation of Cheops' pyramid.

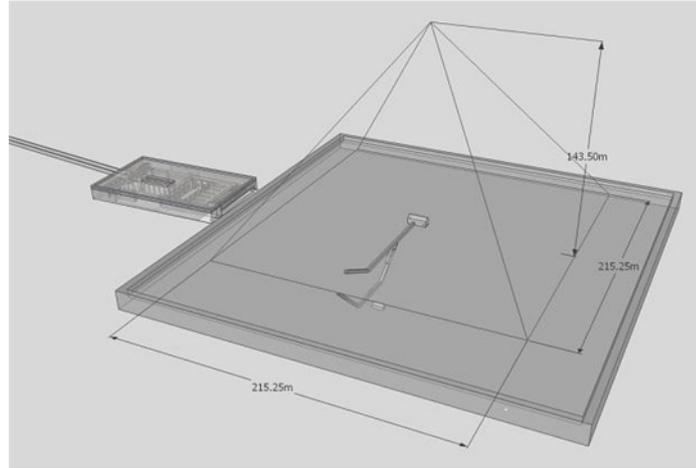

Figure 2: 3D model of Chephrens' pyramid.

Taking into account this displacement, we calculate the distance covered by muons, which reach the chamber, passing through the walls of the pyramid. As an assumption, we consider only muons that enter the pyramid from the top. We used the equation of the plane surface N, W, S and E of the pyramid, equation (1), to calculate the flux of the muons.

$$A = a\sqrt{\left(\frac{b}{2}\right)^2 + h^2} + b\sqrt{\left(\frac{a}{2}\right)^2 + h^2} \qquad (1)$$

With *a* the base length, *b* the base width and *h* the pyramid height.

We introduced the scale factor "*k*" due to the symmetry:

$$k = \frac{143.5}{4} \qquad (2)$$

In the mathematical model developed by us of the Chephren pyramid, we assumed that the pyramid is a regular pyramid with a quadratic base. The height of the pyramid: *h* = 4*k*, the side of the base: *a* = 6*k*. Height of the sidewall of the pyramid: *l* = 5*k* (Egyptian triangle).

From equations (1) and (2) we derive the formulas for the distances from any point on the side wall to the Belzoni Chamber.
The distance *t* to the Belzoni Chamber from the East lateral face (Figure 3, Figure 4) is:

$$t = \frac{8y_0 + 24}{3\sqrt{3} - 4\sin a} \qquad [m] \qquad (3)$$

Where $x_0, y_0$ are the parameters that introduce the shift of the centre of the chamber with respect to the centre of the pyramid, and $\alpha$ is the azimuthal angle.

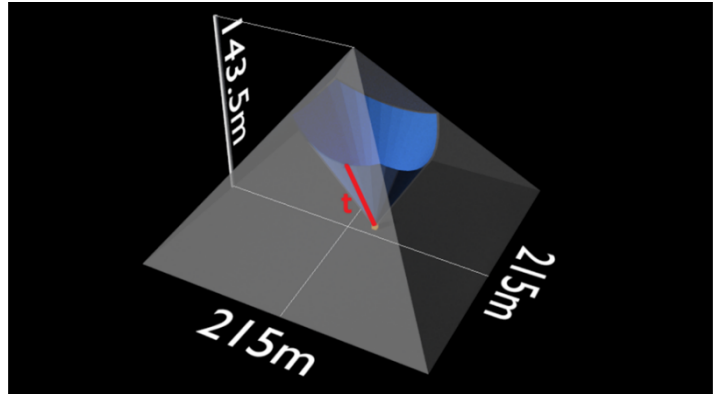
Figure 3: Graphic representation of our mathematical model.

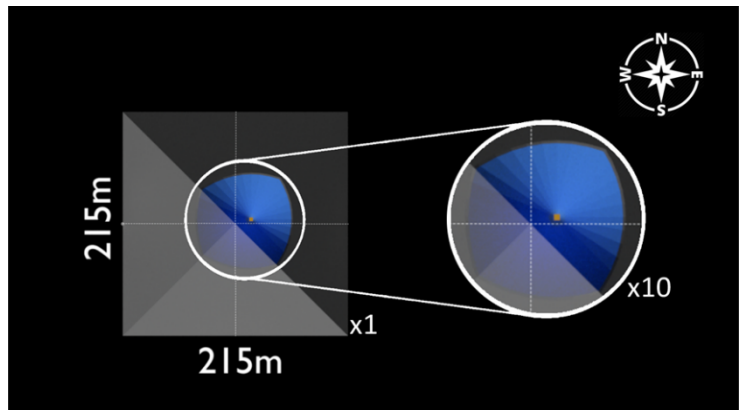
Figure 4: Graphic representation of our mathematical model (vertical projection). The yellow spot represents the location of the Belzoni Chamber.

## Cosmic muons

The integral sea-level muon spectra can be approximated by a power law [6]:

$$N(E_\mu, R) = A E_\mu^{-\gamma} \qquad (4)$$

Where $N$ is the integral energy spectrum of muons, $E_\mu$ is the energy of muons in GeV, $\gamma$ is the exponent of the energy spectrum of muons, $R$ is the range of the muons inside the limestone rock (that the pyramid is made of) and $A$ is a normalisation factor.

Assuming that the muon energy losses increase linearly with their energy $E_\mu$, we derived the formula for the range $R$ of the muons:

$$R = \frac{1}{b} \ln\left(1 + \frac{b}{a} E_\mu\right) \qquad (5)$$

Where $a$ describes the average ionization energy loss in the energy range relevant to the problem and $b$ summarizes processes of muon Bremsstrahlung, direct electron pair creation and nuclear interaction.

From eq. (5) we have obtained the formula for the energy of muons as a function of the constants $a$, $b$ and the range of muons $R$:

$$E_\mu = \frac{a}{b}\left(e^{Rb} - 1\right) \qquad (6)$$

From Taylor's theorem we know that:

$$e^{Rb} \approx 1 + Rb \qquad (7)$$

By substituting (6) into (7) we get:

$$E_\mu = \frac{a}{b}\left(e^{Rb} - 1\right) \approx \frac{a}{b}(1 + Rb - 1) \qquad (8)$$

Hence:

$$E_\mu \approx aR \qquad (9)$$

The range of muons inside the rock can be written as:

$$R = \rho t \qquad (10)$$

Where $\rho$ is the rock density in g/cm³, and $t$ is the distance in cm.

By substituting eq. (9) and (10) into eq. (4) we get the number of muons that reach the detectors inside the Belzoni chamber:

$$N(E_\mu, R) = A(a\rho t)^{-\gamma} \qquad (11)$$

By using eq. (11) for two different distances $t_1$ and $t_2$, we can determine the ratio of the number of muons registered for the two considered distances:

$$\frac{N(t_1)}{N(t_2)} = \frac{A(a\rho t_1)^{-\gamma}}{A(a\rho t_2)^{-\gamma}} = \left(\frac{t_1}{t_2}\right)^{-\gamma} \qquad (11)$$

Using the above equations we estimate the number of muons that reach the chamber, from θ=30° zenith angle and different azimuth angles. Our results, together with the data from [5] can be seen in Figure 5. We observe an unexpectedly high peak for θ=270° (perhaps there is a corridor or irregularities in the construction). Between angles θ=62° and θ=72° we see a possibly higher number of events in comparison to the model. A possible explanation could be that a chamber is located there, which is similar in size to the King Chamber in the Great Pyramid. Another area of interest is between θ=192° and θ=203°. An object of much higher density (e.g. pharaoh gold) could explain this reduced number of muons.

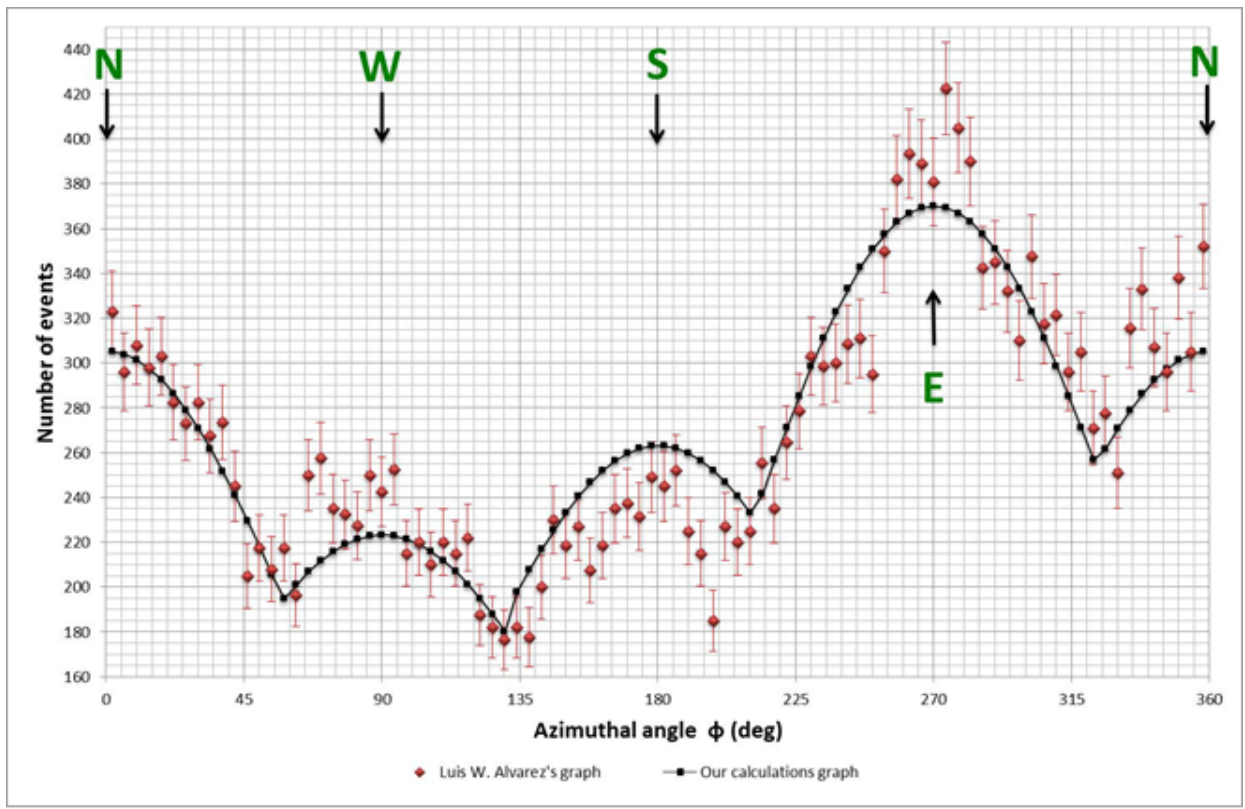

Figure 5: Results for 650.000 insident muons from zenith angles between 20° and 40°. Our calculations are normalized to the number of muons (180) registered for the 130 degree angle.

## The experiment at Beamline for Schools

To increase our understanding on the interactions of muons with matter we submitted our written proposal [7] to the Beamline for Schools competition of 2016 [8] at CERN. More specifically our aim was to understand how the thickness of the limestone walls of the pyramids affects the number of muons that penetrate them, for a given energy.

### Beam characteristics

The incoming proton beam from the Proton Synchrotron accelerator impinges on the primary target and produces secondary particles, which enter the T9 experimental area [9]. The collisions of the protons with the target provide a variety of particles, such as electrons, positrons, pions, kaons and (anti)protons. Muons are also present in the beam, from the decay of pions and kaons. The T9 beam line, used for the experiment, is therefore a mixed hadron and electron beam and can transport either positively or negatively charged particles with momenta between 0.5 GeV/c and 10 GeV/c. The beam arrives in bursts of 0.4 seconds. Depending on the scheduling, such a burst is provided typically once or twice per minute. The maximum number of particles per burst of $10^6$ is achieved for a 10 GeV/c positive beam, but drops for lower momenta. For a negative beam, the rates are typically lower. The beam travels approximately 55 m before it enters the experimental area. The fixed setup of detectors consist of two Cherenkov counters, one Scintillator and one Delay Wire Chamber (DWC) (Figure 6).

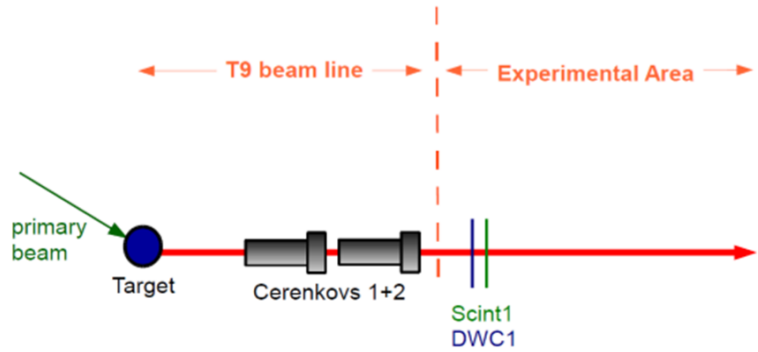

Figure 6: Sketch of the beamline in the T9 experimental area.

## Experimental setup

A schematic drawing of our experimental setup is shown in Figure 7 and a panoramic photo in Figure 8. Two iron blocks of total length of 160 cm at the entrance of T9 to serve as muon filters. The muon beam exiting the iron block is defocused and has a wide momentum distribution. The muon momentum is reduced by an average of approximately 1 GeV/c for 80 cm of iron. Two scintillators in coincidence, after the muon filter, with a distance of 2.7 m between them count the number of muons that will enter the limestone. The limestone was positioned on a movable table. After the limestone, another scintillator coincidence measures the muons that passed through the entire width of the limestone.

The area of coincidence of the detectors is 20x24 cm for the detectors SC1, SC2 and SC3, SC4. The choice of the area of coincidence for both systems ensured the maximum acceptance of focused muons that pass through the entire width of limestone.

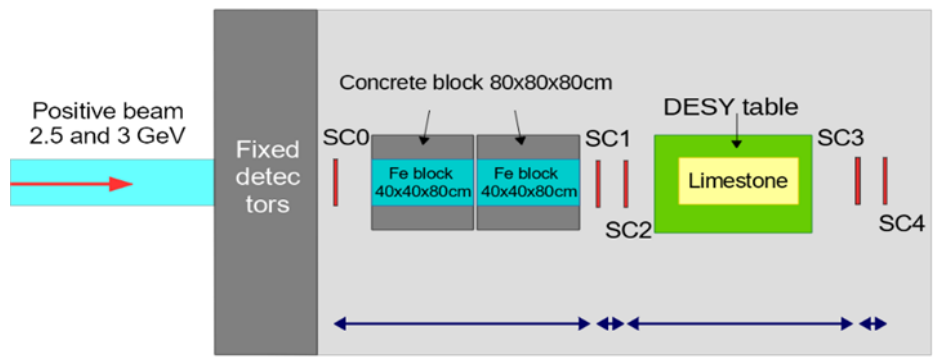

Figure 7: Schematic drawing of the experimental setup.

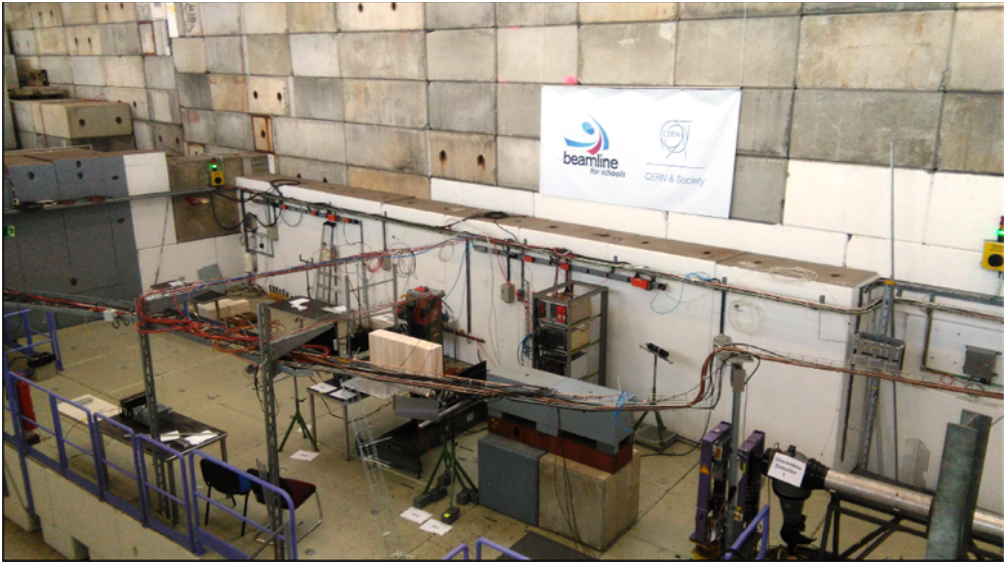

Figure 8: Panoramic photo of the experiment in the T9 hall.

## Results

We first measured the number of muons that passed through the coincidences of the detectors SC1, SC2 and SC3, SC4 without limestone (reference measurement). Then we took measurements placing 10 cm thick limestone blocks (10 x 20 x 40 cm) between them. The total thickness of the limestone ranged from 10 cm to 140 cm. The measurement was repeated for beam momenta of 2.5 GeV/c and 3.0 GeV/c. The ratio of muons, which passed through the limestone over the number of muons recorded by both coincidences, without the limestone was calculated for the different lengths of blocks of limestone. A plot of this ratio with respect to the limestone thickness can be seen in Figure 9 for 2.5 GeV/c and Figure 10 for 3 GeV/c. The red curve corresponds to the best fit to the experimental data (eq. (12) and (13) respectively).

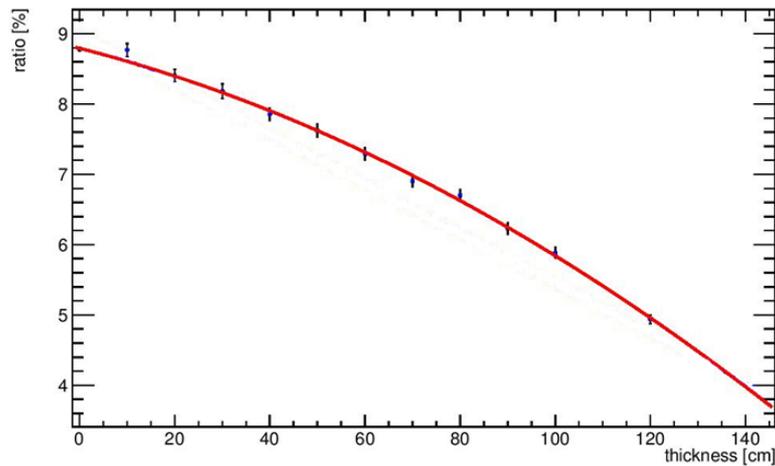

Figure 9: Ratio of muons that passed through the limestone over the initial number of muons, as a function of thickness of limestone for 2.5 GeV/c.

$$y = (-1.386 \cdot 10^{-4} \pm 0.001 \cdot 10^{-4})x^2 - (1.772 \cdot 10^{-2} \pm 0.002 \cdot 10^{-2})x + 8.8507 \pm 6 \cdot 10^{-4} \quad (12)$$

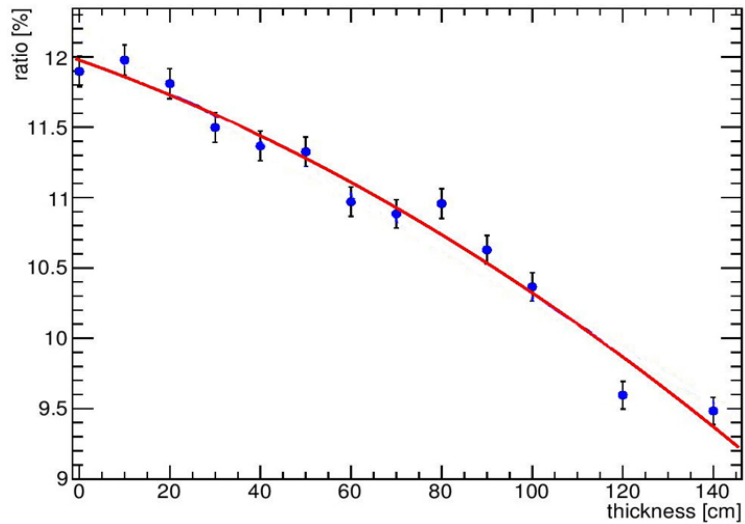

Figure 10: Ratio of muons that passed through the limestone over the initial number of muons, as a function of thickness of limestone for 3 GeV/c.

$$y = (-6.74 \cdot 10^{-5} \pm 0.01 \cdot 10^{-5})x^2 - (9.19 \cdot 10^{-3} \pm 0.02 \cdot 10^{-3})x + 11.909 \pm 6 \cdot 10^{-4} \quad (13)$$

## Muon tomography

The above measurements served as a calibration to our muon tomography setup. To prove the validity of the fits of eq. (12) and (13) the staff of Beamline for Schools placed an unknown structure on the motorised table of known external dimensions (Figure 11). Its external dimensions were 60cm x 60 cm and consisted of 20cm x 20cm squares with or without limestone. We performed a 2D scan of the structure, using a primary beam of 2.5 GeV/c, by moving it by 20 cm at a time. In total six measurements were taken.

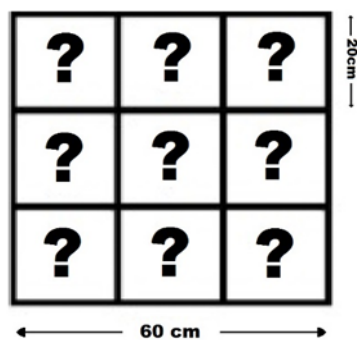

Figure 11: A limestone structure of unknown inner shape was positioned on the motorized „DESY" table. It was scanned with the muon beam both in X and Y direction.

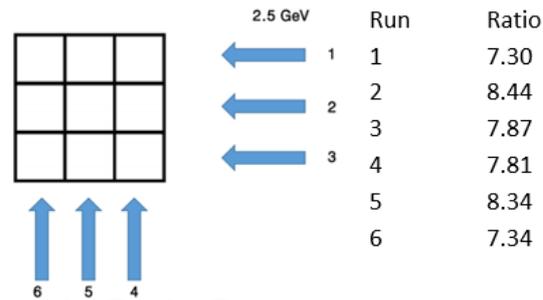

Figure 12: We performed a two-dimensional scan of the unknown structure, which resulted in 6 measurements.

By using eq. (12) we computed the thickness of the limestone structure for each direction and we recontructed the structure. The uncertainty of our prediction of the was around 2 cm.

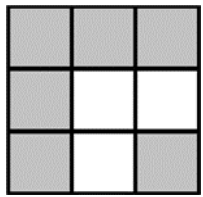

Figure 13: The reconstructed limestone structure. The grey squares correspond to limestone bricks and the white to empty space.

# Conclusions

With our measurements we performed and verified a calibration of the attenuation of muons for primary beam momenta of 2.5 GeV/c and 3 GeV/c at the T9 experimental area of CERN. The obtained results gave us an insight on how the thickness of a limestone block affects the flux of muons. Our goal is to use these results together with Geant 4 simulations to try to assess the role of fluctuations in the range of muons in the limestone rocks and to accurately calculate the intensity of cosmic ray muons in the Belzoni chamber of the Chephren pyramid.

# Acknowlegements

The authors would like to thank the BL4S support scientists, in particular Theodoros Vafeiadis and Oskar Wyszynski, for their help with the execution of the experiment and the writing of this paper.

The underlying experiment was supported by the CERN & Society Foundation, funded in part by the Arconic Foundation, with additional contributions from the Motorola Solutions Foundation, the Ernest Solvay Fund managed by the King Baudouin Foundation, as well as National Instruments.